\documentclass[fleqn,usenatbib]{mnras}
\usepackage{newtxtext,newtxmath}
\usepackage[T1]{fontenc}
\usepackage{ae,aecompl}
\usepackage{graphicx}% Include figure files
\usepackage{dcolumn}% Align table columns on decimal point
\usepackage{bm}% bold math
\usepackage{color}
\usepackage{natbib}
\usepackage{multirow}

\usepackage{xspace}
\usepackage{nicefrac}
\newcommand{\KEPLER}{{\textsc{Kepler}}\xspace}

\newcommand{\B}[2]{\ensuremath{[\text{#1}/\text{#2}]}\xspace}
\newcommand{\Msun}{\ensuremath{\mathrm{M}_\odot}}

\newcommand{\Ep}[1]{\ensuremath{10^{#1}}}
\newcommand{\E}[1]{\ensuremath{\times\Ep{#1}}}

\newcommand{\erg}{\ensuremath{\mathrm{erg}}}

\title{On the Core-Collapse Supernova Explanation for LAMOST J1010+2358}

\author[S. K. Jeena et al.]{
S. K. Jeena,$^{1}$\thanks{E-mail: jeenaunni44@gmail.com} Projjwal Banerjee$^{1}$, and Alexander Heger$^{2,3,4}$
\\
$^{1}$Department of Physics,  Indian Institute of Technology Palakkad, Kerala, India\\
$^{2}$School of Physics and Astronomy, Monash University, Vic 3800, Australia\\
$^{3}$OzGrav: The ARC Centre of Excellence for Gravitational Wave Discovery, Australia\\
$^{4}$ARC Centre of Excellence for Astrophysics in Three Dimensions (ASTRO-3D), Australia\\
}

\date{Accepted XXX. Received YYY; in original form ZZZ}

\pubyear{2023}
\begin{document}
\label{firstpage}
\pagerange{\pageref{firstpage}--\pageref{lastpage}}
\maketitle
\begin{abstract}
Low-metallicity very massive stars with an initial mass of $\sim 140$--$260\,\Msun$ are expected to end their lives as pair-instability supernovae (PISNe). The abundance pattern resulting from a PISN differs drastically from regular core-collapse supernova (CCSN) models and is expected to be seen in very metal-poor (VMP) stars of ${\rm[Fe/H]}\lesssim -2$.  Despite the routine discovery of many VMP stars, the unique abundance pattern expected from PISNe has not been unambiguously detected. The recently discovered VMP star LAMOST J1010+2358, however, shows a peculiar abundance pattern that is remarkably well fit by a PISN, indicating the potential first discovery of a bonafide star born from gas polluted by a PISN.  In this paper, we study the detailed nucleosynthesis in a large set of models of CCSN of Pop III and Pop II star of metallicity ${\rm[Fe/H]}=-3$ with masses ranging from $12$--$30\,\Msun$.  We find that the observed abundance pattern in LAMOST J1010+2358 can be fit at least equally well by CCSN models of $\sim 12$--$14\,\Msun$ that undergo negligible fallback following the explosion. The best-fit CCSN models provide a fit that is even marginally better than the best-fit PISN model.  We conclude the measured abundance pattern in LAMOST J1010+2358 could have originated from a CCSN and therefore cannot be unambiguously identified with a PISN given the set of elements measured in it to date.  We identify key elements that need to be measured in future detections in stars like LAMOST J1010+2358 that can differentiate between CCSN and PISN origin.
\end{abstract}


\section{Introduction}
Very metal-poor (VMP) stars with $\B{Fe}{H}\leq -2$ are crucial for exploring the chemical evolution of the early Galaxy within the first Gyr from the Big Bang.   Until that time, core-collapse supernovae (CCSNe) resulting from the death of massive stars of $\gtrsim 8\,\Msun$ are the dominant contributors to nucleosynthesis. In particular, the surface composition of low mass VMP stars of $\lesssim 0.8\,\Msun$ are fossil records of the composition of the interstellar medium (ISM) present in the early Galaxy.  Current chemical evolution models suggest that many VMP stars have formed from a gas cloud predominantly polluted by the explosion of a single massive star \citep{ryan1996,ritter2012,chiaki2018}. 
This, in turn, can be used to infer detailed information about the nucleosynthesis from individual massive stars, which can be used to gain insight into their masses and the associated initial mass function (IMF) of first-generation (Pop III) and early (Pop II) massive stars. We therefore expect to observe the abundance pattern that results from the explosion of very massive stars of $\sim 140$--$260\,\Msun$ that end their life as pair-instability supernovae (PISNe) \citep{ober1983,heger2002} in at least some of the low-mass VMP stars.  In particular, for Pop III stars, simulations suggest a top-heavy IMF with many very massive stars \citep{abel1998first, abel2002formation, Hirano2015MNRAS}. 
The abundance patterns produced by Pop III PISNe, however, are markedly different from those produced by CCSNe that result from typical massive stars with initial masses of $\lesssim 100\,\Msun$ \citep{heger2002}. 
PISNe produce abundance patterns that have a large deficit of odd $Z$ elements such as Na, Al, P, Cl, and K relative to even $Z$ elements when compared to regular CCSN. 
Thus, the abundance pattern in VMP stars formed from gas polluted by a single PISN should be easily identifiable.  Despite the discovery of several hundreds of VMP stars with sufficiently detailed abundance patterns and the claim of a potential VMP star with PISN signature \citep{aoki2014}, until recently, no clear candidate VMP stars have been identified that show a clear signature arising from a PISN. 

This situation, however, has changed with the recent discovery of the VMP star LAMOST J1010+2358 (hereafter J1010+2358) by \citet{xing2023Natur}.  This star has a peculiar abundance pattern that has been shown to be well fit by a PISN resulting from a Pop III star of $260\,\Msun$ with a He core of $130\,\Msun$. The peculiar features observed in the abundance pattern of J1010+2358 that make this star stand out compared to other VMP stars in the halo are the very low upper limit of Na abundance of $\B{Na}{Fe}<-2.02$ along with the highly sub-solar value of Mg of $\B{Mg}{Fe}=-0.66$ as well as a sub-solar value of $\B{Ca}{Fe}=-0.13$. The sub-solar value of $\alpha$ elements such as Mg and Ca is usually attributed to contributions from SN 1a~\citep{iwamoto1999,ohshiro2021ApJ}.  This star, however, also has sub-solar values [X/Fe] for elements from Ca to Zn and, in particular, of $\B{Cr}{Fe}$ and $\B{Mn}{Fe}$.  \citet{xing2023Natur} found that this pattern was incompatible with an abundance pattern arising from the mixture of SN 1a and CCSN yields, leaving PISN origin as the most likely explanation.  \citet{xing2023Natur} also explored a large range of "classical" massive star CCSN models from the literature, however, none of these models provided a better match. 

In this paper, we calculate the nucleosynthesis in massive stars of initial mass ranging from $12$--$30\,\Msun$ that undergo CCSN with a standard explosion energy of $1.2\E{51}\,\erg$, with primordial (Pop III) and $[Z]=-3$ (Pop II) initial composition.  We find that the observed abundance pattern of J1010+2358 can be fit remarkably well using regular CCSN models provided they do not undergo fallback of material containing Fe group elements.  The quality of fit is even better or at least comparable to those found by \citet{xing2023Natur} using a $260\,\Msun$  of \citet{heger2002}.   

The layout of the paper is as follows: In Section~\ref{sec:method}, we briefly describe the methods used for the models.  The details of the evolution and nucleosynthesis in CCSN models and the best-fit models for J1010+2358, and the comparison with PISN models are discussed in Section~\ref{sec:results_discussion}. Finally, we conclude with a summary of the paper in Section~\ref{sec:conclusion}.

\section{Methods}\label{sec:method}
We simulate the evolution and nucleosynthesis of non-rotating stars of initial mass ranging from $12\,\Msun$ to $30\,\Msun$, with an initial composition corresponding to the primordial Big Bang nucleosynthesis that is adopted from \citet{cyburt2002}.  In the mass range of $12$--$15\,\Msun$ we use intervals of $0.1\,\Msun$, for $15$--$20\,\Msun$ we use $0.2\,\Msun$ intervals, and for $20$--$30\,\Msun$ we use $0.5\,\Msun$ intervals.  We designate models with primordial metallicity as \texttt{z} models and label the models with their initial mass. For example a \texttt{z} model of $12\,\Msun$ is referred to as \texttt{z12}.
We also simulate models on the same mass grid but with an initial metallicity of $\Ep{-3}$ of the solar metallicity where we use the abundances from Big Bang nucleosynthesis for elements up to Li and scaled solar abundances from \citet{asplund2009} for all elements from Be to Zn.  We refer to these as the \texttt{v} models which are also labelled with their initial progenitor mass.
We use the 1D hydrodynamic stellar evolution code \KEPLER \citep{weaver1978presupernova,rauscher2003hydrostatic} to follow the evolution of the star from its birth to its death via CCSN and calculate the detailed nucleosynthesis using a large adaptive co-processing network with reaction rates based on \citet{rauscher2002nucleosynthesis}. The explosion is modelled by a spherically symmetric piston starting from the base of the oxygen shell that coincides with the radius where the entropy per baryon exceeds $4\,k_{\rm B}$ similar to earlier studies such as \citet{heger2010nucleosynthesis}. We label the mass coordinate corresponding to this radius as $M_{\rm cut, ini}$, which we assume to collapse and form the proton-neutron star.

\section{Results and Discussion}\label{sec:results_discussion}
The evolution of single massive stars and the resulting nucleosynthesis has been studied extensively over the last several decades and our current understanding is discussed in detail in the review by \citet{woosley+2002}.  Broadly speaking, for Pop III and Pop II stars of low metallicity, stars of progenitor mass $\sim 10$--$30\,\Msun$ undergo collapse of the central Fe core that can, in many cases, result in a successful explosion via the neutrino-driven mechanism, leading to a CCSN that leaves behind either a neutron star or a black hole.  Stars of $30$--$100\,\Msun$ also undergo core collapse but are unlikely to undergo a successful explosion and collapse into a black hole~\citep{muller2016simple}.  Stars of $\sim 100$--$140\,\Msun$ have He cores of $\sim 40$--$64\,\Msun$ result in instability caused by electron-positron pair creation at the core and are referred to as pulsation PISN (PPISN).  Such stars undergo several pulsations following central C burning that can eject the entire H envelope and even some of the material from the He shell.  Stars in this mass range eventually undergo core collapse to form a black hole. Stars of $140$--$260\,\Msun$ have He cores of $\sim 64$--$130\,\Msun$ undergo a single pulse due to pair-instability following central C burning that completely disrupts the stars leading to an energetic explosion resulting in a PISN. 

Below, we briefly summarise some of the key features of nucleosynthesis in non-rotating Pop III and Pop II massive stars of initial mass $12$--$30\,\Msun$ that result in CCSN with a typical explosion energy of $\sim \Ep{51}\,\erg$.  We focus on the major isotopes of key elements.  The purpose of this review is to put in context the nucleosynthesis site for which later ejection or fallback determines the resulting abundance patterns.

\begin{figure}
    \centering
    \includegraphics[width=\columnwidth]{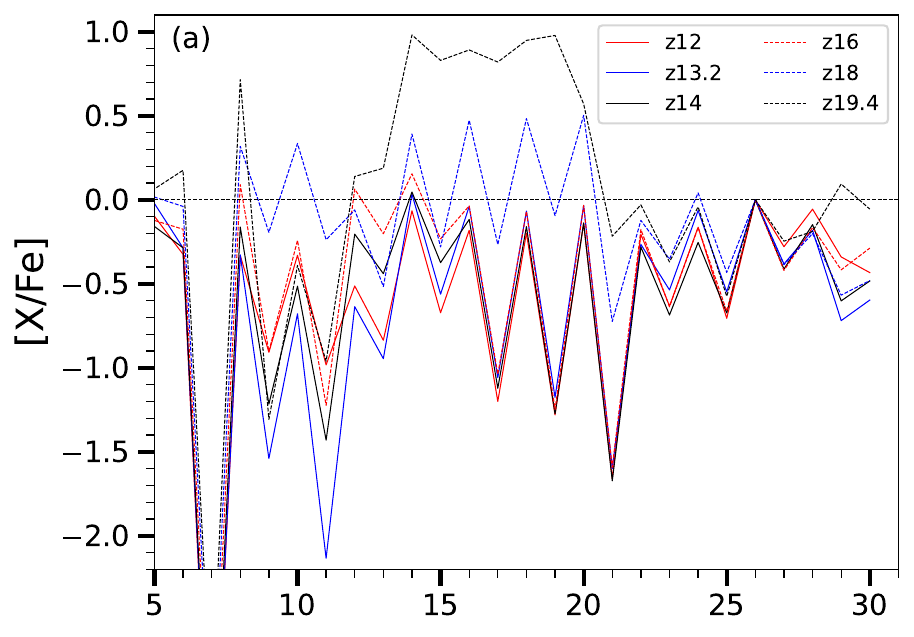}
    \includegraphics[width=\columnwidth]{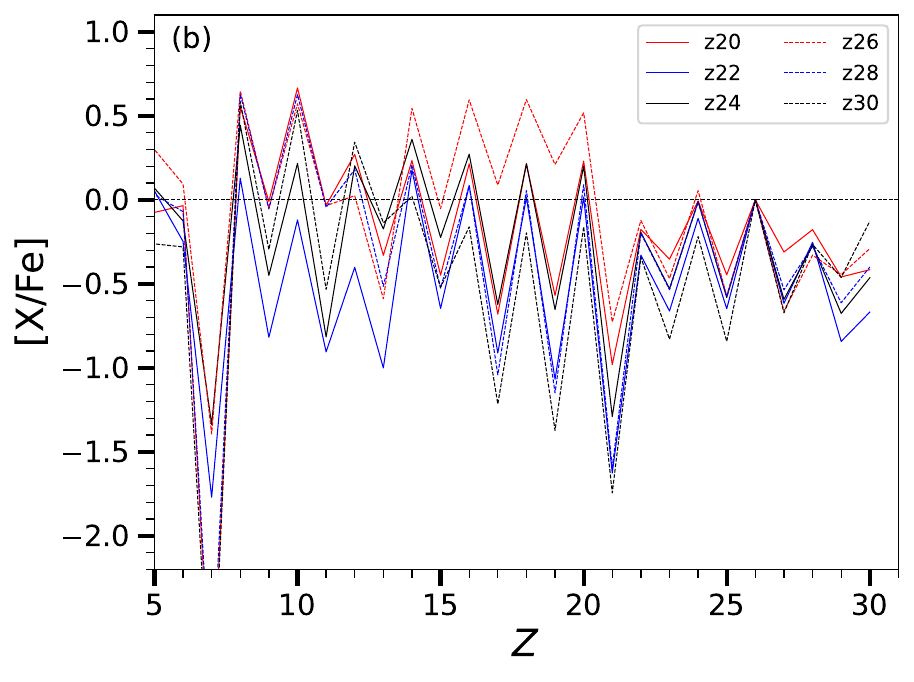}
    \caption{Elemental pattern relative to Fe from the CCSN ejecta for \texttt{z} models of $12$--$30\,\Msun$ in steps of 2 for standard SN explosion energy of $1.2\E{51}\,\erg$ without mixing and fallback.}
    \label{fig:ccsn_zmodel_nofallback}
\end{figure}

\subsection{Nucleosynthesis up to core collapse}

A massive star first undergoes core H burning followed by core He burning. The primary product following core He burning is $^{12}$C and $^{16}$O. The next burning stage is core C burning which primarily produces $^{20}$Ne, $^{24}$Mg, $^{23}$Na and $^{27\!}$Al. Next  $^{20}$Ne is burned via $(\gamma,\alpha)$,  leaving behind $^{16}$O, $^{24}$Mg, and $^{28}$Si as the main product. During this stage, the $\alpha$ particles released via $(\gamma,\alpha)$ also burn $^{23}$Na to $^{27\!}$Al. 
Following $^{20}$Ne depletion, the core contracts and as the temperature reaches $\sim 2\E{9}\,$K, core O burning ignites resulting in  $^{28}$Si and $^{32}$S as the main products.  During early oxygen burning as the core grows into the O-Ne-Mg shell that surrounds it, the Mg in the core is destroyed as well.
As the core contracts even further, core Si burning first results in a quasi-equilibrium of isotopes of several $\alpha$ elements and odd $Z$ elements. When the core contracts further, nuclear statistical equilibrium (NSE) is established that is dominated by iron group isotopes which form the Fe core that ultimately collapses into either a neutron star or a black hole. Material outside the Fe core comprises concentric shells of progressively lighter elements that are left behind in earlier burning stages. In addition, while the core is burning heavier fuel in the centre, partial or even complete shell burning of lighter fuel also takes place.  This is particularly relevant during the final phase of the star's life after core O depletion.  At that stage, a star may exhibit concurrent shell burning of C, Ne, and even O. In particular, convective \emph{shell} Ne burning usually leads to the destruction of $^{23}$Na via $^{23}$Na$(\alpha,\gamma)^{27\!}$Al.  If, however, the convective Ne shell grows and mixes in material from the C-O shell, then the destruction of Na can be mitigated by Na production due to C burning.  Shell O burning can be an important source of large amounts of isotopes of Si to Sc.   Otherwise, these are created mostly by explosive O burning during the CCSN \citep{ritter2018}.  If, however, shell O burning takes place before the collapse, it can produce isotopes of Si to Sc in amounts comparable to or in excess of what is produced during the explosion.

\begin{figure}
    \centering
    \includegraphics[width=\columnwidth]{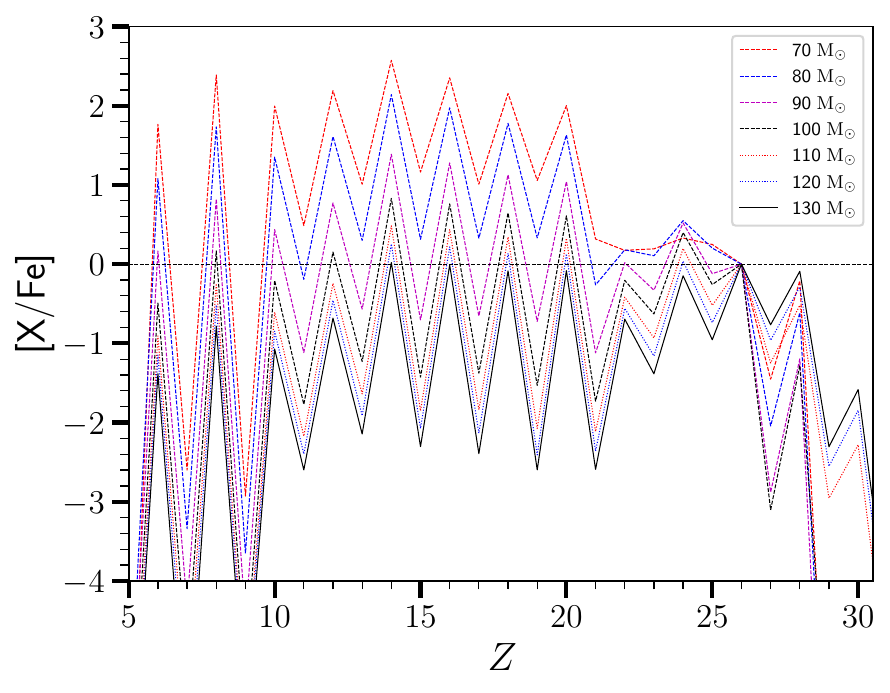}
    \caption{Elemental pattern relative to Fe resulting from PISN models from \citet{heger2002} that are labelled with the corresponding He core mass ranging from $70$--$130\,\Msun$. Data is adapted from \textsc{StarFit} \citep{heger2010nucleosynthesis}.}
    \label{fig:PISN}
\end{figure}

In some cases, the convective O-burning shell can merge with the convective O-Ne-Mg shell, resulting in a mixing of products of O burning further out into the star, which can result in a large enhancement of isotopes of Si to Sc in the final ejecta and reduced the amplitude of odd-even abundance pattern.
The mixing, however, can also lead to the destruction of Na because the material of the O-Ne-Mg shell is mixed into the hotter regions of the O-burning shell. In many cases we have studied, however, the combined convectively-mixed O-burning and O-Ne-Mg shell grows and mixes in material from the C-O shell, which can mitigate the destruction of Na. In this case, the destruction of Na depends sensitively on the temperature at the base of the convective O burning shell.

\subsection{Explosive nucleosynthesis following core collapse}
%When the mass of the iron core exceeds the Chandrasekhar mass, the core collapses into a proto-neutron star which results in a bounce along with neutrino emission of $\Ep{57}$ neutrinos on timescales of seconds. With the help of the neutrino, the shock wave propagates outwards and ejects the material above the iron core. 
Explosive nucleosynthesis occurs when the CCSN shock travels through the mantle. The high temperature in the post-shock region processes the innermost parts of the ejecta, which comprises the O-Si shell into Fe peak elements ranging from Ti to Zn via complete O and Si burning. The outer regions of the O-Si and O-Ne-Mg shells undergo incomplete Si burning as well as O burning that primarily produces isotopes of elements from Si to Sc. Further out, the shock can also burn some of the Ne in the O-Ne-Mg shell resulting in the usual Ne burning product, i.e., $^{16}$O and $^{24}$Mg. Notably, similar to Ne shell burning during the pre-SN stage, explosive Ne burning destroys $^{23}$Na. The enormous amount of neutrinos emitted by the proto-neutron star can also lead to neutrino-induced spallation reactions ($\nu$-process) leading to substantial production of $^{7}$Li, $^{11}$B, and $^{19}$F \citep{woosley1990,heghax2005}.

\subsection{Abundance Pattern from CCSN}
The final abundance pattern emerging from the ejecta of a single CCSN depends not only on the details of nucleosynthesis but also on the mixing of material from different parts of the ejected core and the amount of material that falls back. An exact calculation of mixing and fallback requires full 3D hydrodynamic simulations that model the explosion using full-neutrino transport in order to get a self-consistent explosion via the neutrino-driven mechanism. Such computations, however, are much too expensive to be employed on a large set of models. Usually, mixing and fallback are treated in 1D explosion models in a parametric fashion by treating them as free parameters to fit an abundance pattern such as \citet{nomoto2013nucleosynthesis,tominaga2014abundance}. Alternatively, fallback from a spherically symmetric explosion can be calculated by studying the long-term behaviour of the ejecta that depends on the explosion energy, which is again a free parameter. Then the mixing is treated in a uniform manner for all models that is only calibrated to fit the light curve for SN 1987A for a specific progenitor \citet{heger2010nucleosynthesis}. 

Both approaches require some amount of fallback of the innermost ejecta that contains the Fe group elements in order to fit the abundance pattern observed in VMP stars, in particular, to match the high $\B{X}{Fe}\gtrsim 0.3$ for the alpha elements. 

\begin{figure}
    \centering
    \includegraphics[width=\columnwidth]{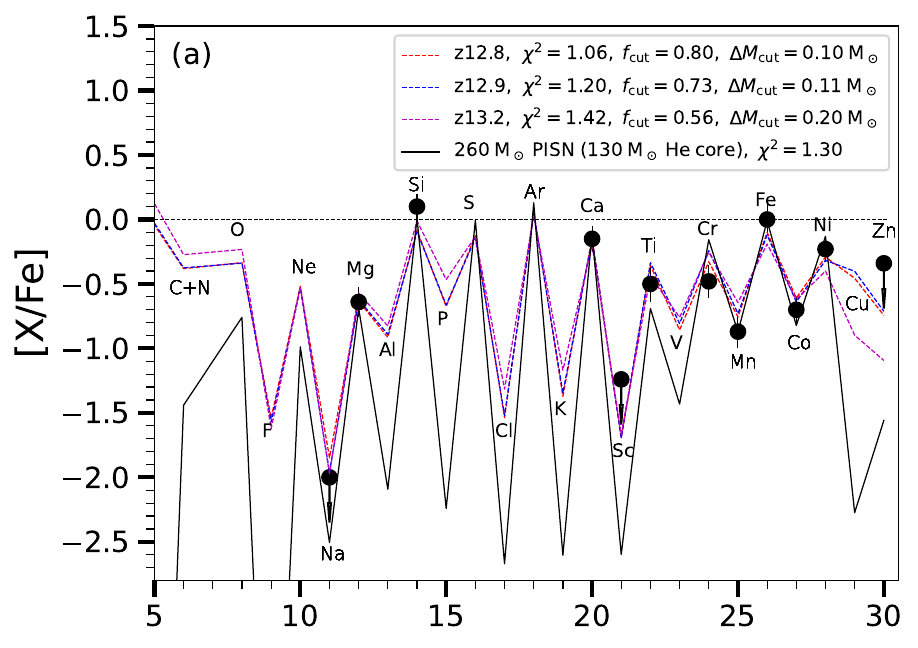}
     \includegraphics[width=\columnwidth]{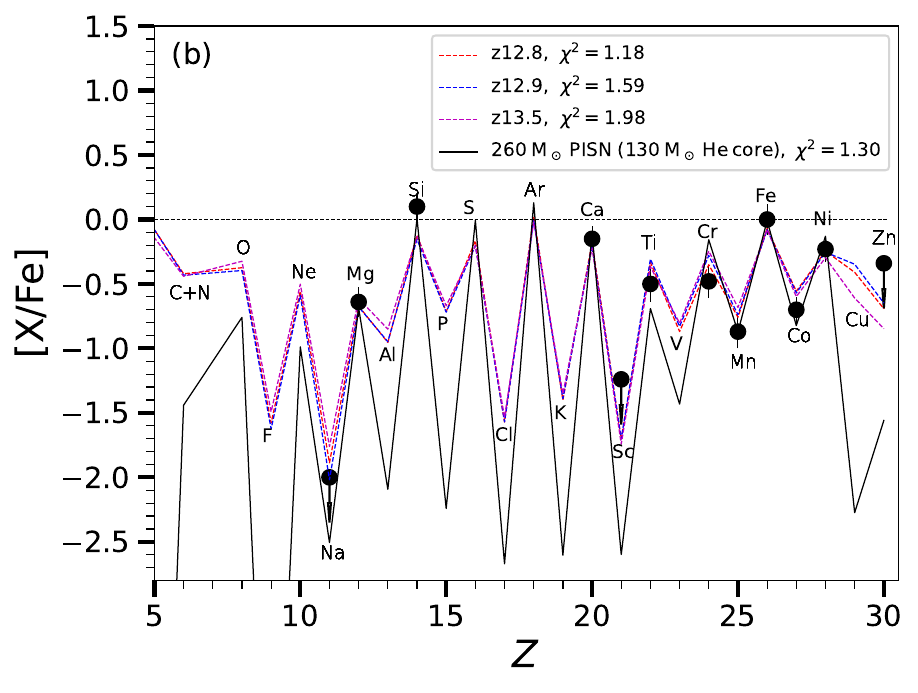}
    \caption{\textbf{(a)} The chemical abundances of J1010+2358 compared with the top three best-fit models from mixing and fallback models of CCSN of $12$--$30\,\Msun$ with explosion energy of $1.2\E{51}\,\erg$ mixed with an ISM of primordial composition.  \textbf{(b)} Same as (a) but without any fallback.}
    \label{fig:standard_EXPLD_z}
\end{figure}

\subsection{CCSN Ejecta from \texttt{z} Models}
It is instructive to look at the abundance pattern resulting from the ejecta without any mixing and fallback.  Figure \ref{fig:ccsn_zmodel_nofallback} shows the abundance pattern from the CCSN ejecta with fiducial explosion energy of $1.2\E{51}\,\erg$ without any fallback for \texttt{z} models ranging from $12$--$30\,\Msun$ where only the models with even integer masses are shown for clarity. In addition, we also plot the \texttt{z13.2} that has one of the lowest levels of \B{Na}{Fe} and \B{Mg}{Fe} and \texttt{z19.4} that undergoes merger of O burning shell with convective O-Ne-Mg shell. The value of \B{X}{Fe} for Fe peak elements from Ti to Zn is sub-solar in all the models and shows a clear odd-even pattern. For $\alpha$ elements from Si to Ca, a large fraction of our models have $\B{X}{Fe}\sim 0$. 
We find that such models do not undergo shell O burning prior to collapse and all of these elements are made predominantly during explosive O burning. On the other hand, models that undergo O shell burning exhibit a large production of elements from Si to Sc,  much more than what is produced during explosive burning. This can be clearly seen in models such as \texttt{z16} and \texttt{z18} that have super-solar values of $\B{X}{Fe}\gtrsim +0.3$ for the $\alpha$ elements from Si to Ca along with enhanced production of odd $Z$ elements from P to Sc. Models that undergo the merger of O shell burning with the convective O-Ne-Mg shell can have a dramatic increase in the abundance of elements from Si to Sc as seen in the \texttt{z19.4} model. These models also have enhanced odd $Z$ elements from P to Sc with a clearly diminished odd-even pattern.

\begin{figure}
    \centering
    \includegraphics[width=\columnwidth]{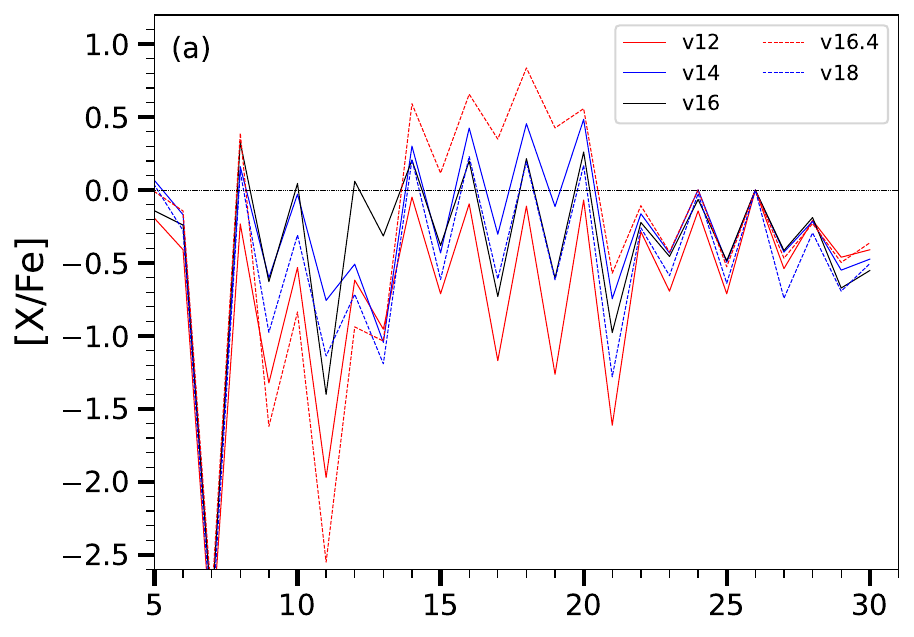}
    \includegraphics[width=\columnwidth]{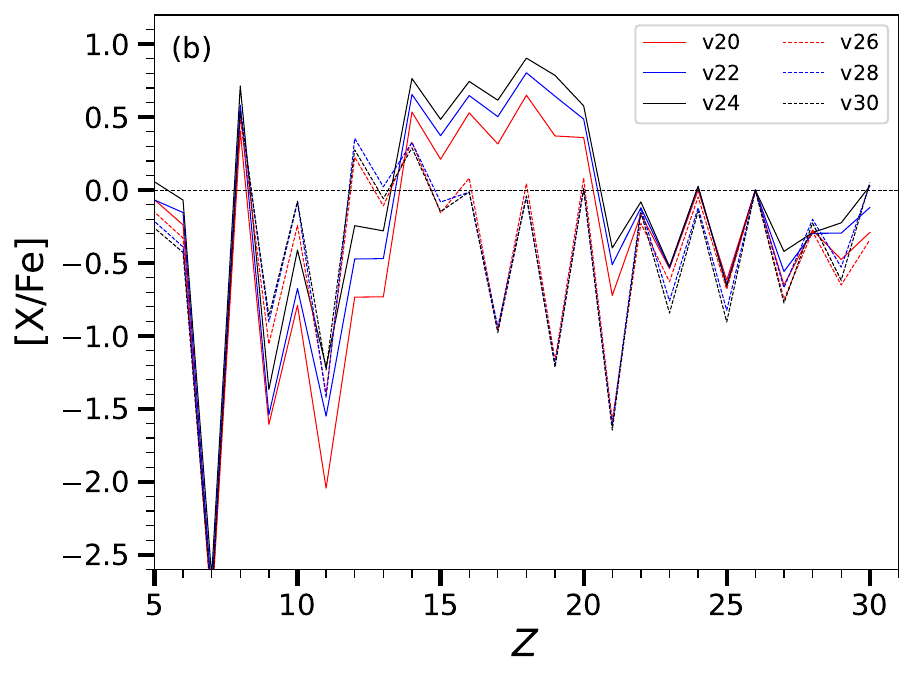}
    \caption{Same as Fig.~\ref{fig:ccsn_zmodel_nofallback} but for \texttt{v} models.}
    \label{fig:ccsn_vmodel_nofallback}
\end{figure}

Mg is distinct from the heavier $\alpha$ elements such as Si and Ca as it is almost exclusively produced during C and Ne burning prior to collapse; explosive burning mostly leads to the destruction of Mg in the innermost part of the O-Ne-Mg shell. The yield of Mg generally scales with the amount of mass in the O-Ne-Mg, which on average increases with progenitor mass. As can be seen from Fig.~\ref{fig:ccsn_zmodel_nofallback}, $\B{Mg}{Fe}\lesssim 0$ from most models with values as low as $-0.7$ for the lower mass models. Only in models $\gtrsim 20\,\Msun$, $\B{Mg}{Fe}>0$ reaches values of up to $\sim +0.3$ due to large O-Ne-Mg shell. The trend of Al is very similar to Mg as it is also created during C and Ne burning and typically has sub-solar values of \B{Al}{Fe} that correlate with \B{Mg}{Fe}.
Ne also behaves similarly to Mg.   Almost all of the ejected Ne is produced during C burning prior to collapse. Thus, \B{Ne}{Fe} typically also has values $\lesssim 0$ for most of the models of $<20\,\Msun$, with values as low as $-0.7$ for lower mass models of $13$--$14\,\Msun$. For models of $\gtrsim 20\,\Msun$, the \B{Ne}{Fe} can reach super-solar values of up to $+0.6$ due to that much larger O-Ne-Mg shells. F is predominantly produced by neutrino spallation on Ne \citep{woosley1990}, which is also partially destroyed by the SN shock heating. Overall, \B{F}{Fe} correlates with \B{Ne}{Fe} and can have values as low as $-1.5$ in lower mass models that have very low [Ne/Fe].

Na is similar to Ne as it is produced during C burning and \B{Na}{Fe} overall correlates with \B{Ne}{Fe} and thus also with \B{F}{Fe}. Na, however, is more susceptible to destruction in the O-Ne-Mg shell when there is Ne shell burning. Even if Ne barely burns, it is sufficient to destroy Na via $^{23}$Na$(\alpha,\gamma)^{27}$Al. This occurs at the final stages of models of $\gtrsim 12.7$--$16.4\,\Msun$ where convective Ne burning just prior to core-collapse burns substantial amounts of Na, leading to $\B{Na}{Fe}\lesssim -1$. \B{Na}{Fe} is lowest in lower mass models of $\sim 13\,\Msun$, where the O-Ne-Mg shell is less massive and a large fraction of the Na is also destroyed by the passage of the supernova shock wave that heats the O-Ne-Mg shell to higher temperatures due to the compact structure of the lower mass models. This can result in \B{Na}{Fe} values as low as $\sim -2.2$. Due to Na burning into Al, lower mass models that have a low \B{Na}{Fe} do not have similarly low values of \B{Al}{Fe}. Interestingly, the low mass models from $12$--$12.6\,\Msun$ do not have a very low \B{Na}{Fe} as they do not undergo convective Ne burning. Furthermore, in these models, the O-Ne-Mg shell extends out to a larger radius where the shock wave becomes less energetic, such that the destruction of Na by shock heating is lower.

\begin{figure}
    \centering
    \includegraphics[width=\columnwidth]{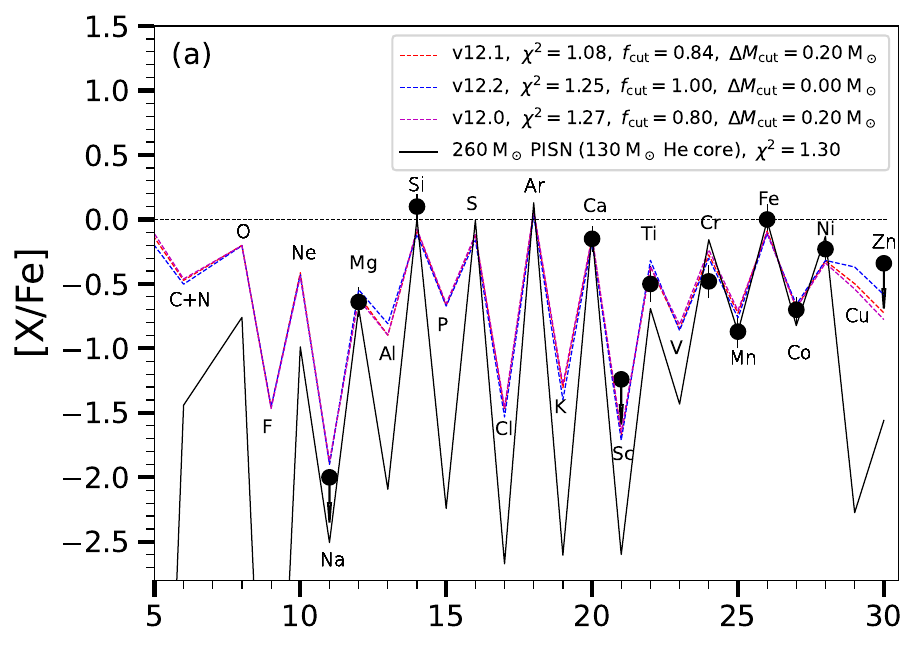}
    \includegraphics[width=\columnwidth]{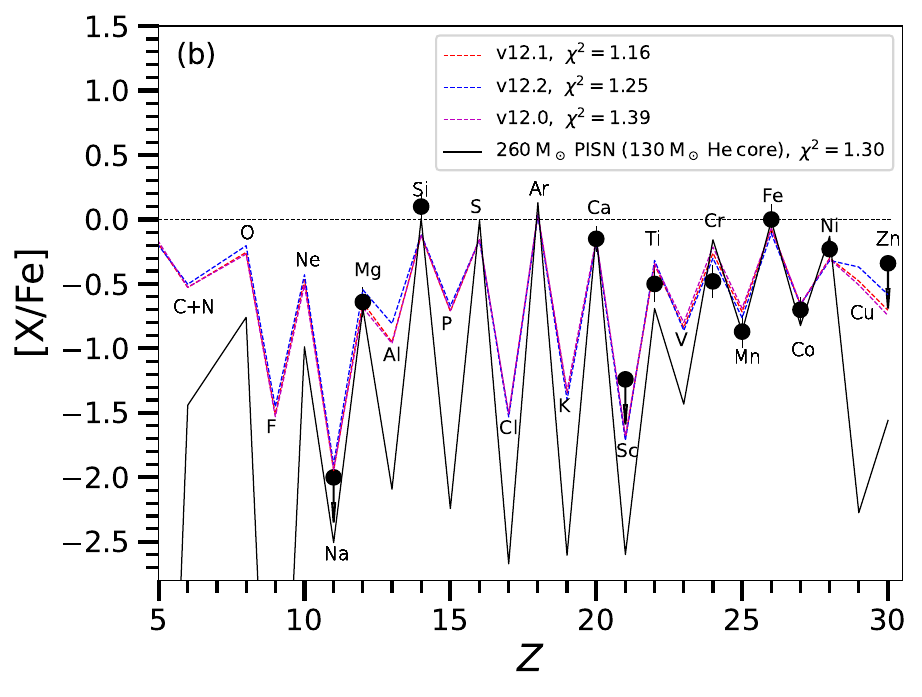}
    \caption{Same as Fig.~\ref{fig:standard_EXPLD_z} but for \texttt{v} models.}
    \label{fig:standard_EXPLD_v}
\end{figure}

\subsection{PISN Yield Pattern}
Considering that \citet{xing2023Natur} found that their best-fit for J1010+2358 was a PISN, we briefly discuss how the PISNe abundance patterns compare to CCSN yields from our \texttt{z} models discussed above. Figure~\ref{fig:PISN} shows the PISN abundance pattern from Pop III stars of He core mass ranging from $70$--$130\,\Msun$ that are adapted from the \textsc{StarFit} database \citep{heger2010nucleosynthesis} based on the calculations by \citet{heger2002}.  These models cover the mass range of stars that can undergo PISN. The figure shows that the PISN yield pattern varies substantially with the mass of the He core. In particular, the yield of Fe peak elements increases rapidly relative to the light and intermediate elements up to Sc. For example, the amount of Fe produced increases from $\sim 0.1\,\Msun$ for the $70\,\Msun$ He core model to $\sim 40\,\Msun$ for the $130\,\Msun$ model \citep{woosley+2002}. This results in a large variation of \B{X}{Fe} for C to Sc ranging from highly super-solar for lower mass models to sub-solar for the heaviest PISN model. In particular, the abundance ratios of $\alpha$ elements from Si to Ca, $\B{X}{Fe}\sim 0$ found in J1010+2358 are only produced in models with He core masses of $\gtrsim 120\,\Msun$. The high-mass models are also the only ones which have large sub-solar $\B{Mg}{Fe}\lesssim -0.5$ along with very low $\B{Na}{Fe}\lesssim -2$ similar to what is found in J1010+2358. Thus, it is clear that only the most massive PISN can match the yield pattern found in J1010+2358. The abundance pattern from the most massive models has similar features to some of the lower mass \texttt{z} models of $\sim 13$--$14\,\Msun$ with respect to the observed elements in J1010+2358. Except for Na, the higher mass PISN models have large differences in odd $Z$ elements such as F, Al, P, Cl, and K compared to the CCSN models. Additionally, C and O are extremely deficient in higher mass PISN models relative to Fe, which is distinct from CCSN models.

\subsection{Best-fit CCSN from \texttt{z} Model for J1010+2358}
Several of our \texttt{z} models of $12$--$30\,\Msun$ without fallback produce abundance pattern with $\B{X}{Fe}\sim 0$ for $\alpha$ elements from Si to Ca, as well as $\B{X}{Fe}<0$ for Fe peak elements. Additionally, $\B{Mg}{Fe}$ and \B{Na}{Fe} can have highly sub-solar values in models of $\sim 13$--$14\,\Msun$. These are very similar to the key features in the abundance pattern seen in J1010+2358. 
In order to find the best-fit model for J1010+2358, however, we need to consider the abundance pattern emerging from the ejecta from all the models from $12$--$30\,\Msun$ including the possibility that the ejecta from CCSN can undergo both mixing and fallback from aspherical explosions. We model this using a prescription similar to \citet{tominaga2007,ishigaki2014,jeenaCEMP2023} where an additional mass cut $M_{\rm cut, fin}$ is introduced above which all material is ejected whereas only a fraction $f_{\rm cut}$ of the mass $\Delta M_{\rm cut}=M_{\rm cut, fin}-M_{\rm cut, ini}$ is ejected. In this case, 
$f_{\rm cut}$ and  $M_{\rm cut, fin}$ are free parameters where the former varies from $0$ to $1$ and we vary $M_{\rm cut, fin }$ in steps of $0.1\,\Msun$ from a minimum value of $M_{\rm cut, ini}$ to a maximum value corresponding to the enclosed mass of the base of the H envelope. We note here that $M_{\rm cut, ini}$ is not a free parameter as it is fixed at the mass coordinate where entropy per baryon exceeds $4 k_{\rm B}$.  The abundance yield, $Y_{{\rm X}_i}$, of any element ${\rm X}_i$, defined as the sum over all isotopes of the ejecta mass fractions divided by their corresponding mass numbers, depends on $M_{\rm cut, fin}$ and $f_{\rm cut}$. Thus, the ratio of the total abundance of any element $X_i$ relative to a reference element $X_{\rm R}$ can be written as 
\begin{equation}
    \frac{ N_{{\rm X}_i}}{ N_{\rm   X_R}}=\frac{ Y_{{\rm X}_i}(M_{\rm cut,fin},f_{\rm cut})}{ Y_{\rm X_R}(M_{\rm cut,fin},f_{\rm cut})}  
\end{equation}
The best-fit model is then found by minimizing deviation from the observed value to the model using a $\chi^2$ prescription that also takes into account the observed uncertainty $\sigma_i$ for each element as described in detail in \citet{heger2010nucleosynthesis} and more recently in \citet{jeenaCEMP2023}. Here we set $\sigma_i=\max(\sigma_i,0.1)$ in order to avoid making $\chi^2$ overly sensitive to elements that have a very low value of $\sigma_i$.

Figure~\ref{fig:standard_EXPLD_z}a shows the top three best-fit \texttt{z} models with standard explosion energy of $1.2\E{51}\,\erg$. We also plot the best-fit PISN model from Pop III star $260\,\Msun$ reported in \citet{xing2023Natur}, where the data is adapted from \textsc{StarFit} database \citep{heger2010nucleosynthesis}. All of the top three best-fit models provide an excellent fit to the observed abundance pattern. The top two best-fit models, \texttt{z12.8} and \texttt{z12.9}, have a slightly lower $\chi^2$ than the best-fit PISN model whereas \texttt{z12.8} has marginally higher $\chi^2$. It is important to note that although the abundance patterns from best-fit CCSN models differ dramatically from the best-fit PISN model, which has a huge deficiency of odd $Z$ elements, for the limited set of elements that are measured in J1010+2358, there are only very minor differences. 
Among these minor differences, $\B{Ti}{Cr}\sim -0.1$ to $-0.2$ in our best-fit CCSN models is more consistent with the observed value of -0.07 compared to $-0.55$ in the PISN model.  \B{Co}{Ni} is about $-0.35$ and $-0.65$ in the CCSN and PISN model, respectively, compared to the observed value of $-0.55$. In this case, values from both models are consistent with the observed value within the observational uncertainties. 
Another difference is the lower $\B{Na}{Fe}$ in the PISN model compared to the CCSN model. Although the observed upper limit is consistent with the best-fit CCSN model, it definitely fits better with the PISN model. We find, however, that the \B{Na}{Fe} in CCSN models can be easily reduced by $0.2$--$0.3$ when the explosion energy is increased slightly to $1.5\E{51}\,\erg$ without affecting the other yields.
We find that in all of the best-fit CCSN models, the amount of matter that undergoes fallback, $(1-f_{\rm cut})\Delta M_{\rm cut}$, is minimal ranging from $0.02-0.09\,\Msun$. Figure~\ref{fig:standard_EXPLD_z}b shows the corresponding top three best-fit models without any fallback. As can be seen from the figure, even without any fallback and the associated adjustable free parameters, the top three models provide an excellent fit with slightly lower or comparable $\chi^2$ compared to the best-fit PISN model. 

\subsection{CCSN Ejecta from \texttt{v} Models and Best-fit for J1010+2358}
Figure~\ref{fig:ccsn_vmodel_nofallback} shows the abundance pattern of selected \texttt{v} models from $12$--$30\,\Msun$ with a standard explosion energy of $1.2\E{51}\,\erg$ without any mixing and fallback. The overall features are very similar to those of the \texttt{z} models discussed earlier. Again, we find $\B{X}{Fe}\lesssim 0$ for Fe peak elements for all models. Also, stars that do not undergo shell O burning have $\B{X}{Fe}\sim 0$ for $\alpha$ elements from Si to Ca, whereas models that do undergo O burning have a large enhancement of elements from Si to Sc.  Interestingly, in the \texttt{v} models, stars such as \texttt{v16.4} that undergo a merger of the O burning shell with the convective O-Ne-Mg shell, which leads to a large destruction of Na that is not seen in the \texttt{z} models. The primary reason for this is that the temperature at the base of the O burning shell in \texttt{v} models is slightly higher than the \texttt{z} models which leads to a large destruction of Na. Such models, however, cannot provide a good fit for J1010+2358 as they have elevated, super-solar abundances of $\alpha$ elements from Si to Ca.

Similar to our \texttt{z} models, lower mass \texttt{v} models have sub-solar \B{X}{Fe} for Mg, Na, Ne, and F. In slight contrast, however, the lowest mass \texttt{v} models starting from $12\,\Msun$ undergo shell Ne burning prior to collapse and have a structure similar to $\sim 13\,\Msun$ \texttt{z} models. Thus, these are again ideal candidates for matching the abundance pattern for J1010+2358. Figure~\ref{fig:standard_EXPLD_v}a  shows the top three best-fit \texttt{v} models compared to the best-fit PISN model from \citet{xing2023Natur}. Again, the three best-fit \texttt{v} models \texttt{v12.1}, \texttt{v12.2}, and \texttt{v12} provide excellent fits to the observed abundance pattern of J1010+2358 with slightly lower $\chi^2$ values compared to the best-fit PISN model. Similar to the \texttt{z} models, the level of fallback in the best-fit models is negligible, with fallback masses ranging from $0$--$0.04\,\Msun$. Figure~\ref{fig:standard_EXPLD_v}b shows the corresponding best-fit without any fallback where, again, the top three best-fit models provide an excellent fit with slightly lower or comparable $\chi^2$ value than the best-fit PISN model. We note here that for the \texttt{v} models, we have assumed a primordial composition for gas in the ISM with which the SN ejecta mixes. This can occur in the early Galaxy where CCSN ejecta can mix inhomogeneously with inflowing primordial gas to form the next generation of stars. On the other hand, if the gas in the ISM has a metallicity corresponding to the initial metallicity of the \texttt{v} models, the final composition of the gas mixed with SN ejecta will not reflect the composition of the SN ejecta. In particular, for elements that have highly subsolar values of [X/Fe] from the SN ejecta such as Na and Mg, the values will be altered significantly and will increase towards solar values of [X/Fe] after mixing with the ISM. In such a case, the unique pattern for J1010+2358 cannot be produced by \texttt{v} models.

\section{Summary and Conclusions}\label{sec:conclusion}
We find that the peculiar abundance pattern observed in J1010+2358 can be very well reproduced by CCSN models $\sim 12$--$13\,\Msun$ of both Pop III \texttt{z} and Pop II \texttt{v} stars. We find that the quality of the best-fit from CCSN models is marginally better than the best-fit PISN model. Remarkably, the best-fit CCSN models are characterised by negligible fallback and the best-fit CCSN models that have no fallback provide equally good fits. Since $M_{\rm cut, ini}$ is not a free parameter in our calculation, the best-fit CCSN models with negligible or no fallback implies that essentially no free parameters such as $M_{\rm cut, fin}$ and $f_{\rm cut}$ are required to fit the abundance pattern.
%Remarkably, the best-fit CCSN models are characterised by negligible fallback and thus have essentially no free parameters to fit the abundance pattern. 
Judging by the quality of fit quantified by the $\chi^2$, both the CCSN models and the PISN models provide an equally good fit for the observed abundance pattern for J1010+2358. 

One of the striking differences between CCSN and PISN is indeed the large deficiency of odd $Z$ elements F, Al, P, Cl, and K that could be used to clearly differentiate the two scenarios. None of these elements, however, are observed or have strong upper limits in J1010+2358. Future detection or strong upper limits on any of these elements will allow us to distinguish between CCSN and PISN models. Among the odd $Z$ elements, Na is a notable and crucial exception, where the large observed deficiency of Na with an upper limit of $\B{Na}{Fe}<-2.02$ is not unique to PISN but can also be found in CCSN models. This is due to the distinct origin of Na, which is produced entirely due to C burning compared to heavier odd $Z$ elements from P to Sc that are produced during O burning.  Moreover, because Na has fewer protons than heavier odd $Z$ elements, it is much more susceptible to being destroyed by CCSN shock heating. 
Among other elements that are clearly different in the best-fit CCSN and PISN models are C and O. Because the best-fit CCSN models do not undergo significant fallback, they have the lowest value form \B{C}{Fe} and \B{O}{Fe} of $\sim -0.5$ that is possible from a CCSN model. The best-fit PISN model, however, has a much lower value of $\B{C}{Fe}=-1.4$ and $\B{O}{Fe}=-0.75$. Thus, a detection or a strong upper limit of C or O could help distinguish between CCSN and PISN as the source of elements in J1010+2358. We note here that the lack of heavy elements found in J1010+2358 as indicated by the strong upper limit of $\B{Sr}{Fe}<-2.5$ and $\B{Ba}{Fe}<-1.17$ cannot be used to distinguish between PISN and CCSN because neither PISN nor our regular Pop III CCSN models produce any heavy elements.  \emph{The observational determination of the key odd-$Z$ elements along with C and O mentioned above} is equally critical to other future or past PISN-origin candidates.

Another important feature of J1010+2358 is its highly sub-solar $\B{Mg}{Fe}=-0.66$. Such low values of Mg are usually not attributed to CCSN, but we find that it is in fact a general feature in many of the models of $\lesssim 20\,\Msun$ provided fallback is negligible. Specifically, as long as the innermost part of the ejecta that contains the Fe peak elements is ejected, CCSN can result in sub-solar values of \B{Mg}{Fe}. Traditionally, Mg-poor VMP stars have been associated with pollution by SN Ia rather than a CCSN \citep{ivans2003ApJ,Li2022ApJ}. Our study indicated that some of the Mg-poor stars could be the result of CCSN that do not undergo substantial fallback of Fe peak elements. We find that the same models that do not undergo fallback and that have sub-solar Mg also have $\B{X}{Fe}\sim 0$ for alpha elements from Si to Ca which is also observed in J1010+2358. In future, we plan to explore other VMP stars that have sub-solar \B{Mg}{Fe} to see if some of them have their origins from CCSN instead of SN 1a. 

The peculiar abundance pattern measured in J1010+2358 compared to other VMP stars measured in the Galactic halo and the rarity of such abundance patterns potentially have important ramifications. On one hand, it could point to the low chance of detecting a star directly formed from a gas polluted by a PISN. This can be due partly to the rarity of the most massive He core progenitors required for PISN to produce the Mg and Na poor pattern observed in J1010+2358.  This is in contrast to the high fraction of low-mass CCSNe progenitors among all stars that make CCSNe in a standard initial mass function, which, though, may not apply to Pop III stars.
Another factor that could reduce the chance of such stars occurring in nature is that the chance of forming a star from a gas polluted exclusively by a PISN is very rare due to the large explosion energies associated with PISN. Such energetic explosions lead to gas outflows from the host minihalo resulting in large-scale dilution and mixing \citep{chiaki2018}. 
On the other hand, if CCSNe are indeed the source of the elements measured in  J1010+2358, the rarity of such patterns implies that CCSN that do not undergo fallback is quite rare, even among low-mass CCSN progenitors. 
This is supported by the fact that the majority of VMP stars measured in the Galactic halo have super-solar values for $\alpha$ elements even though many CCSN without fallback can result in subsolar \B{X}{Fe} for Mg and solar values for Si and Ca. In fact, in previous studies by \citet{heger2010nucleosynthesis} and \citep{tominaga2007}, substantial fallback of the innermost ejecta containing the Fe peak elements was usually adopted in the 1D CCSN mixing and fallback models in order to match the abundance pattern in VMP stars. The level of fallback in the best-fit CCSN models of $12$--$14\,\Msun$ and the fraction of models that do not undergo fallback can only be clarified by a larger set of 3D simulations of CCSN explosion in future.

\section*{Data Availability}

Data is available upon reasonable request.

\section*{Acknowledgements}
This work was supported by the Science and Engineering Research Board Grant no SRG/2021/000673.
A.H.\ was supported by the Australian Research Council (ARC) Centre of Excellence (CoE) for Gravitational Wave Discovery (OzGrav) through project number CE170100004, by the ARC CoE for All Sky Astrophysics in 3 Dimensions (ASTRO 3D) through project number CE170100013, and by ARC LIEF grants LE200100012 and LE230100063.

\bibliography{reference}
\bibliographystyle{mnras}

\end{document}